\documentclass[doublecol]{epl2} 

\usepackage{amssymb}
\usepackage{amsmath}

\title{Non-perturbative renormalization-group approach to zero-temperature
  Bose systems}
  
\author{N. Dupuis\inst{1,2,3} \and K. Sengupta\inst{4}}
\shortauthor{N. Dupuis \etal}

\institute{                    
  \inst{1} Laboratoire de Physique Th\'eorique de la Mati\`ere Condens\'ee, 
CNRS - UMR 7600, \\ Universit\'e Pierre et Marie Curie, 4 Place Jussieu, 
75252 Paris Cedex 05,  France 

\inst{2} Laboratoire de Physique des Solides, CNRS - UMR 8502,
  Universit\'e Paris-Sud, 91405 Orsay, France

\inst{3} Department of Mathematics, Imperial College, 
180 Queen's Gate, London SW7 2AZ, United Kingdom

\inst{4} TCMP division, Saha Institute of Nuclear
Physics, 1/AF Bidhannagar, Kolkata-700064, India
}
\pacs{05.30.JP}{Boson systems}
\pacs{03.75.Hh}{Static properties of condensates}
\pacs{05.10.Cc}{Renormalization group methods}

\abstract{We use a non-perturbative renormalization-group technique to study
  interacting bosons at zero temperature. Our approach reveals the
  instability of the Bogoliubov fixed point when $d\leq 3$ and yields the exact
  infrared behavior in all dimensions $d>1$ within a rather simple theoretical
  framework. It also enables to compute the low-energy properties in terms of
  the parameters of a microscopic model. In one-dimension and for not
  too strong interactions, it yields a good picture of the Luttinger-liquid
  behavior of the superfluid phase.}

\begin{document}

\newcommand{\mean}[1]{\langle #1 \rangle}
\newcommand{\cc}{{\rm c.c.}} 
\newcommand{\Tr}{{\rm Tr}} 
\def\half{\frac{1}{2}}  
\def\dt{\partial_t}
\def\w{\omega} 
\def\q{{\bf q}} 
\def\r{{\bf r}} 
\def\eps{\epsilon} 
\def\nablabf{\boldsymbol{\nabla}}
\def\beq{\begin{equation}}
\def\eeq{\end{equation}}
\def\bleq{\begin{eqnarray}}
\def\eleq{\end{eqnarray}} 
\def\inttau{\int_0^\beta d\tau}
\def\inttaur{\int_0^\beta d\tau \int d^dr}
\def\dtau{{\partial_\tau}} 

\maketitle

\section{Introduction}

In spite of the success of the Bogoliubov theory in providing a microscopic
explanation of superfluidity \cite{Bogoliubov47}, a clear understanding of the
infrared behavior of interacting boson systems at zero temperature has
remained a challenging theoretical issue until very recently. Besides
approximations that do not satisfy the Goldstone-Hugenholtz-Pines theorem
\cite{Hohenberg65,Hugenholtz59}, first attempts to improve 
the Bogoliubov theory revealed a singular perturbation theory plagued
by infrared divergences due to the presence of the Bose-Einstein condensate
and the Goldstone mode \cite{Beliaev58b,Gavoret64}. These
divergences cancel in most physical quantities but lead to a vanishing of
the anomalous self-energy $\Sigma_{\rm an}(q)$ in the limit $q\equiv(\q,\w)\to
0$ although the linear spectrum and therefore the superfluidity are preserved
\cite{Nepomnyashchii75,Nepomnyashchii78,Nepomnyashchii83,Popov79}. 
This observation seriously  called into question the validity of the
Bogoliubov theory, where the linear spectrum relies on a finite value of
$\Sigma_{\rm an}(q\to 0)$ \cite{note2}. The physical origin
of the vanishing of the anomalous self-energy is the divergence of the
longitudinal correlation function which is driven by the gapless (transverse)
Goldstone mode -- a general phenomenon in systems with a continuous broken
symmetry \cite{Patasinskij74}. 

The infrared behavior of zero-temperature Bose systems is now well understood
in the modern language of renormalization group (RG)
\cite{Weichman88,Benfatto94,Castellani97,Pistolesi04}. 
Using a field-theoretical renormalization-group approach supplemented by the
Ward identities associated with the gauge symmetry, Castellani {\it et
al.} were able to establish the exact infrared behavior of a zero-temperature
Bose system \cite{Castellani97,Pistolesi04}. Only for $d>3$ does the
Bogoliubov theory predict the correct infrared behavior, whereas the
Bogoliubov fixed point is found to be unstable for $d\leq 3$ even though the
low-energy mode remains phonon-like with a linear spectrum. 
In the approach of Refs.~\cite{Castellani97,Pistolesi04}, the low-energy
behavior of the correlation functions is expressed exactly in terms of 
thermodynamics quantities such as the density, the condensate density or the
macroscopic sound velocity. Despite its very elegant formulation, this
approach however does not appear to enable an explicit calculation of the
correlation functions in terms of the parameters of a particular microscopic 
model. Given the present possibilities to realize low-dimensional and/or
strongly correlated Bose systems in ultracold atomic gases \cite{Bloch07}, 
it would be of great interest to have a theoretical framework allowing for
quantitative predictions that could be tested against the experimental
results.  

In this Letter, we use a non-perturbative renor\-mali\-zation-group (NPRG)
technique \cite{Wetterich91,Wetterich93,Berges00,Delamotte07} to
study interacting bosons at 
zero temperature. Not only does our approach give the exact 
asymptotic behavior of the correlation functions in dimensions $d>1$ within a
rather simple theoretical framework free of infrared divergences, but it also
enables to explicitely follow 
the behavior of the system from microscopic to macroscopic scales. In
one-dimension and for not too strong interactions, it yields a good picture of
the Luttinger-liquid behavior of the superfluid phase. NPRG studies of
interacting bosons have previously been reported both at finite
\cite{Andersen99} and zero \cite{Wetterich07} temperature. To a large
extent, our results are complementary to those of Ref.~\cite{Wetterich07}.

\section{Non-perturbative RG approach}

We consider the following action,
\beq
S = \int dx \left[ \psi^*(x)\left(\dtau-\mu - \frac{\nablabf^2}{2m}
  \right) \psi(x) + \frac{g}{2} |\psi(x)|^4 \right] ,
\label{action} 
\eeq
where $\psi(x)$ is a bosonic (complex) field, $x=(\r,\tau)$, $\int
dx=\inttaur$. $\tau\in [0,\beta]$ is an imaginary time, $\beta\to\infty$
the inverse temperature, and $\mu$ denotes the
chemical potential. The interaction is assumed to be local in space and the
model is regularized by a momentum cutoff $|\q|<\Lambda$ (with
$\Lambda\to\infty$ whenever convenient). We take $\hbar=k_B=1$ throughout the
Letter.  

The basic quantity of interest in the NPRG is the effective action
$\Gamma[\phi]$, which is the generating functional of the one-particle
irreducible (1PI) vertices. It is obtained 
by a Legendre transform of the free energy $\ln Z[J]$ computed in the presence
of an external source term $S_J=\int dx [J^*(x)\psi(x) + \cc]$
($\phi=\mean{\psi}_J$) \cite{Berges00,Wetterich07}. To
implement the RG procedure, we add to the action an infrared regulator $\Delta
S_R = \int dx\, \psi^*(x) R(x-x') \psi(x')$ which suppresses the fluctuations
with $\q^2<k^2$. The functional $\Gamma[\phi]\equiv\Gamma_k[\phi]$ then
becomes $k$ dependent and satisfies the exact flow equation
\beq
\dt \Gamma[\phi] = \half \Tr \left\lbrace \dt
R\bigl(\Gamma^{(2)}[\phi]+R\bigr)^{-1}\right\rbrace ,
\label{exact} 
\eeq
where we have introduced the flow parameter $t=\ln(k/\Lambda)$. In Fourier
space, the trace in (\ref{exact}) involves a sum over frequencies and momenta,
as well as a trace over the two indices of the complex field $\phi$.  
$\Gamma^{(2)}[\phi]$ 
is the second-order functional derivative of $\Gamma[\phi]$ with respect to
$\phi$. Choosing $R$ to diverge for $k\to\infty$, all fluctuations are then
suppressed and the mean-field theory, where the effective action
$\Gamma[\phi]$ reduces to the microscopic action $S[\phi]$, becomes
exact. Quantum fluctuations are gradually taken into account by 
decreasing $k$ and making use of (\ref{exact}). For $k=0$, $\Gamma[\phi]$
corresponds to the effective action of the original model (\ref{action})
from which we can deduce all 1PI vertices -- and in particular the
single-particle propagator $G=-\Gamma^{(2)-1}$ -- as well as the thermodynamic
potential. 

The functional differential equation (\ref{exact}) is too complicated to be
solved exactly. For approximate solutions it is sufficient to truncate the
most general form of $\Gamma[\phi]$ \cite{Berges00,Wetterich07}. For a
superfluid Bose system, the simplest choice reads 
\begin{multline}
\Gamma[\phi] = \Gamma_{\rm min} + \int dx \Bigl\lbrace ZZ_1 \phi^*(x) \dtau
\phi(x) - V \phi^*(x) \partial_\tau^2 \phi(x) \\ - Z \phi^*(x)
\frac{\nablabf^2}{2m} \phi(x) + \frac{\lambda}{2} [n(x)-n_0]^2 \Bigr\rbrace ,
\label{ansatz}
\end{multline}
where $n(x)=|\phi(x)|^2$ is the density. 
Eq.~(\ref{ansatz}) is obtained from an expansion to fourth order about the
minimum $|\phi(x)|=\sqrt{n_0}$, where $n_0$ denotes the condensate
density. We use a derivative expansion to order ${\cal O}(\partial^2)$
\cite{Berges00,note1}. For $k\to\infty$, the initial conditions are $Z=Z_1=1$,
$V=0$, $\lambda=g$ and $n_0=\mu/g$, and the effective action $\Gamma[\phi]$
reproduces the Bogoliubov theory. Although $V$ is not present in the original
action (\ref{action}), it is always generated by the flow equation
(\ref{exact}) \cite{Wetterich07} and plays a crucial role when $d\leq 
3$. The degeneracy of the minimum $|\phi(x)|=\sqrt{n_0}$ reflects the
gauge invariance (i.e. the U(1) symmetry $\psi^{(*)}(x)\to\psi^{(*)}(x)e^{\pm
  i\alpha}$) of the action (\ref{action}). A
broken-symmetry state can be obtained by picking up a particular minimum. It is
convenient to write $\phi=(\phi_1+i\phi_2)/\sqrt{2}$ in terms of two real
fields $\phi_1$ and $\phi_2$ and to consider the state
$\bar\phi=(\sqrt{2n_0},0)$ as an example of broken-symmetry state. In Fourier
space, the corresponding single-particle vertex $\bar\Gamma^{(2)}=-\bar G^{-1}$
(with $\bar G_{ij}=-\mean{\psi_i\psi_j}$) then reads
\beq
\bar\Gamma^{(2)}(q) = \left(
\begin{array}{lr}
V\w^2 + Z \eps_\q + 2 \lambda n_0 & ZZ_1\w \\
-ZZ_1\w & V\w^2 + Z \eps_\q 
\end{array}
\right) 
\label{gamma2} 
\eeq 
($\eps_\q=\q^2/(2m)$). 
The vanishing of $\bar\Gamma_{22}(q=0)$, which is a mere consequence of the
U(1) symmetry, naturally implements the Hugenholtz-Pines theorem
\cite{Hugenholtz59,note3} in our formalism. The combination 
$\lambda n_0$ corresponds to the anomalous self-energy $\Sigma_{\rm an}(q=0)$.
There are two important quantities that can be read off from (\ref{gamma2}),
namely the superfluid density $n_s$ and the Goldstone mode velocity $c$,
\beq
n_s = Z n_0 , \quad c = \left( \frac{Z/2m}{V+(ZZ_1)^2/(2\lambda n_0)}
\right)^{1/2}. 
\eeq
For $k\to\infty$, one has $n_s=n_0=\mu/g$ and $c=\sqrt{n_0g/m}$.
The superfluid density is defined in the usual way from the stiffness of the
system with respect to a twist of the phase of the superfluid order parameter
$\bar\phi=(\sqrt{2n_0},0)$. The expression of the velocity $c$ follows from
the equation ${\rm det}\,\bar\Gamma^{(2)}(q)=0$ in the limit $q\to 0$. 

Inserting (\ref{ansatz}) into (\ref{exact}), we obtain the flow equations
\begin{multline}
\dt \tilde n_0 =  -(d+\eta+\eta_1) \tilde n_0 \\ \shoveright{ + 16 s \int_\w
\frac{A^2+MA+M^2-\w^2}{D^2} ,   } \\
\shoveleft{ \dt \tilde\lambda
= (d-2+2\eta+\eta_1) \tilde\lambda - 16 s
 \tilde\lambda^2 } \\  \shoveright{ \times
\int_\w \frac{-5A^3 -3MA^2+A[11\w^2-6M^2]+ 7M\w^2-4M^3}{D^3} ,  } \\
\eta =  16 \frac{v_d}{d} \tilde \lambda M \int_\w \frac{1}{D^2} ,  
\hspace{4.35cm} 
\end{multline}
\begin{multline}
\eta_1 = -\eta -16 s
\tilde\lambda^2 \tilde n_0 \int_\w \biggl\lbrace \frac{1}{D^2}  \\
- \frac{(A+B)(3A-B-4Z_2\w^2)}{D^3} \biggr\rbrace , \\
\shoveleft{\dt Z_2 = (2+\eta+2\eta_1) Z_2 - 16 s \tilde\lambda^2\tilde
  n_0 \int_\w \biggl\lbrace \frac{-Z_2}{D^2} }  
\\  
+ \frac{2(A+B)(Z_2B+1)+4Z_2\w^2[Z_2(3A+5B)+2]}{D^3} \\ \shoveright{
- \frac{6\w^2(A+B)[Z_2(A+B)+1](2Z_2B+1)}{D^4} \biggr\rbrace ,  }
\\ 
\dt \tilde\Omega = -(d+2+\eta_1)\tilde\Omega + 8 s \int_\w
  \frac{A+M}{D} , \hspace{1.7cm} 
\label{rgeq} 
\end{multline}
where $A=1+Z_2\w^2$, $B=A+2M$, $D=AB+\w^2$, $M=\tilde\lambda\tilde n_0$,
$v_d^{-1}=2^{d+1}\pi^{d/2}\Gamma(d/2)$, $s=(v_d/d)[1-\eta/(d+2)]$,
and $\int_\w = \int d\w/(2\pi)$. We have
introduced the dimensionless quantities 
\bleq
\tilde n_0 = Z Z_1 k^{-d} n_0 , &&
\tilde\lambda =  Z^{-2} Z_1^{-1} k^d \eps_k^{-1} \lambda , \nonumber \\ 
ZZ_1^2Z_2 = \eps_k V , && \tilde\Omega = Z_1 k^{-d} \eps_k^{-1} \Omega ,
\eleq 
as well as $\eta = -\dt \ln Z$ and $\eta_1 = -\dt \ln Z_1$, 
and chosen the regulator $R(\q^2) = Z(\eps_k-\eps_\q)\theta(\eps_k-\eps_\q)$
\cite{Litim00}. $\Omega$ denotes the thermodynamic potential per unit volume
in the broken-symmetry state $\bar\phi=(\sqrt{2n_0},0)$. 
For $Z_2=0$ (i.e. $V=0$), the integrals over $\w$ can be carried out and we
reproduce the flow equations derived in Ref.~\cite{Wetterich07}. However, the
approximation $V=0$ -- or a mere perturbative treatment of $V$ -- cannot be
used for $d\leq 3$ as it predicts the wrong 
exponent ($2\eps$ instead of $\eps=3-d$) for the divergence of the longitudinal
correlation function, the wrong lower critical dimension (2 instead of 1), and
-- for $V=0$ -- an infinite velocity for the Goldstone mode.

\section{Superfluidity with BEC ($d>1$)} 

\begin{figure}[!ht]
\centerline{\includegraphics[width=6.5cm,clip]{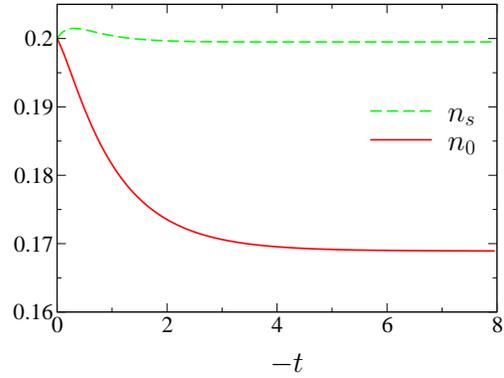}}
\caption{(Color online) Condensate density $n_0$ and superfluid density $n_s$ 
  {\it vs.} $-t$ for $d=2$, $n_0(t=0)=0.2$ and $\lambda(t=0)=10$. Here
  and in the following figures, we use units where $\Lambda=1$ and $2m=1$.}
\label{flow2D_1}
\end{figure}

\begin{figure}[!ht]
\centerline{\includegraphics[width=6.5cm,clip]{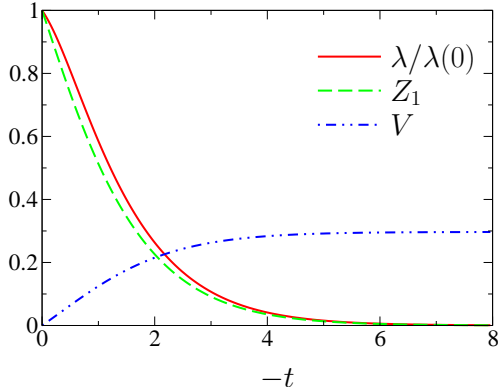}}
\caption{(Color online) $\lambda$, $Z_1$ and $V$ {\it vs.} $-t$ for $d=2$,
  $n_0(t=0)=0.2$ and $\lambda(t=0)=10$.}
\label{flow2D_2}
\end{figure}

When $d>1$, superfluidity is always accompanied by Bose-Einstein condensation
(BEC): $\lim_{k\to 0} n_0=n_0^*>0$. For $d>3$, the Bogoliubov fixed point is
stable; all parameters in the effective action $\Gamma_{t=0}$ remain
finite as $k\to 0$. $V^*=\lim_{k\to 0}V$, although nonzero, gives only a finite
correction to the infrared limit of the vertices. This picture changes
dramatically when $d\leq 3$. In this case, both $Z_1$ and $\lambda$ are
suppressed as $k\to 0$, which explains why the anomalous self-energy
$\Sigma_{\rm an}(q=0)=\lambda n_0$ vanishes in the infrared limit. This
suppression is logarithmic for $d=3$ and powerlaw-like in lower dimensions.
When $d>1$, we can use the fact that $\lim_{k\to 0}M=\lim_{k\to 0}Z_2M=\infty$
to analytically obtain the asymptotic behavior for $k\to 0$ (table
\ref{table}). A typical RG flow in two dimensions is shown in
Figs.~\ref{flow2D_1} and \ref{flow2D_2}. 

\begin{table}[ht!]
\renewcommand{\arraystretch}{1.5}
\begin{center}
\begin{tabular}{|c||c|c|c|}
\hline 
& $d=3$ & $1<d<3$ & $d=1$  
\\ \hline \hline 
$n_0$ & $n_0^*$ & $n_0^*$ & $k^{\eta^*}$ 
\\ \hline 
$n_s$ & $n_s^*$ & $n_s^*$ & $n_s^*$
\\ \hline 
$\lambda$ & $(\ln k)^{-1}$ & $k^\eps$ & $k^{2-2\eta^*}$ 
\\ \hline 
$V$ & $V^*$ & $V^*$ & $k^{-\eta^*}$ 
\\ \hline
$\tilde n_0$ & $k^{-3}/\ln k$ & $k^{2\eps-3}$ & $k^{-\eta_1^*-1}$ 
\\ \hline
$\tilde\lambda$ & $k$ & $k^{1-\eps}$ & $k^{\eta_1^*+1}$
\\ \hline
$\eta$ & $k^2$ & $k^{d-1}$ & $\eta^*$ 
\\ \hline
$\eta_1$ & $Z_1\sim (\ln k)^{-1}$ & $-\eps$ & $\eta_1^*$
\\ \hline
$\eta_2$ & $Z_2\sim (k\ln k)^2$ & $2\eps-2$ & $\eta_2^*=-2\eta_1^*-2$ 
\\ \hline
$\tilde n_0'$ & $k^{-2}$ & $k^{\eps-2}$ & $\tilde n_0'{}^*$ 
\\ \hline 
$\tilde \lambda'$ & $(\ln k)^{-1}$ & $\tilde \lambda'{}^*$ & $\tilde
\lambda'{}^*$ 
\\ \hline 
\end{tabular}
\end{center}
\caption{Asymptotic behavior for $k\to 0$ ($\eps=3-d$). The stared quantities
  indicate nonzero fixed-point values. For $d>1$, these results are obtained
  analytically from the flow equations (\ref{rgeq}). For $d=1$ one
  obtains approximate fixed points rather than true fixed points (see
  text). } 
\label{table}
\end{table}

The suppression of $Z_1$, together with a finite
$V^*$, shows that the effective action exhibits a space-time SO($d$+1) symmetry
in the infrared limit \cite{Wetterich07}. This limit is well understood and
corresponds to the classical O(2) model in $d+1$ dimensions. The symmetry can
be made explicit by the rescaling $\tilde\r=k\r$,
$\tilde\tau=(Z_1\eps_k^{-1}\sqrt{Z_2})^{-1}\tau$ and $\tilde\phi(\tilde
x)=(ZZ_1\sqrt{Z_2}k^{-d})^{1/2}\phi(x)$, whereby the effective action becomes
\begin{multline}
\Gamma[\tilde\phi] = \Gamma_{\rm min} + \int d\tilde x \Bigl\lbrace 
ZZ_2^{-1/2}\eps_k
  \tilde\phi^*(\tilde x) \partial_{\tilde\tau} \tilde\phi(\tilde x) \\ -
  \tilde\phi^*(\tilde x) (\partial^2_{\tilde\tau}+\nablabf^2_{\tilde\r})
  \tilde \phi(\tilde x)  + \frac{\tilde\lambda'}{2}[\tilde n(\tilde x)-\tilde
    n_0']^2 \Bigr\rbrace , 
\label{SO}  
\end{multline}
where $\tilde n_0'=\sqrt{Z_2}\tilde n_0$ and $\tilde\lambda'= \tilde\lambda/
\sqrt{Z_2}$. For a typical frequency $\tilde\w\sim k$, the term 
linear in $\partial_{\tilde\tau}$ becomes subleading with respect to the
quadratic one. Equivalently, one can observe that for $d\leq 3$ the Goldstone
mode velocity reaches the fixed point value $c^*=(Z^*/2mV^*)^{1/2}$ which is
independent of $Z_1$. Our numerical results for the scaling of $\tilde n_0'$
and $\tilde\lambda'$ agree with the known results for the Goldstone regime of
the classical O(2) model in $d+1$ dimensions (table \ref{table}). The
dimensionless coupling $\tilde\lambda'$ vanishes for $d=3$ and flows to a
finite value for $d<3$. The relation 
$2\eta_1^*+\eta_2^*=-2$ (with $\eta_2=-\partial_t \ln Z_2$) ensures that the
Goldstone mode velocity reaches a finite value for $k\to 0$, i.e. that the
dynamical exponent takes the value $z=1$. 

\begin{figure}[!ht]
\centerline{\includegraphics[width=5.25cm,clip,angle=-90,bb=175 475 455 810]{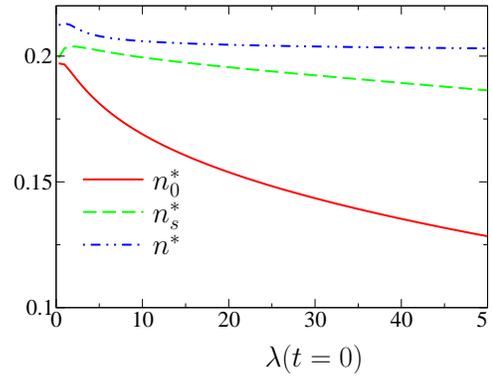}}
\caption{(Color online) $n_0^*$, $n_s^*$ and $n^*$ {\it vs.} $\lambda(t=0)$
  for $d=2$ and $n_0(t=0)=0.2$.} 
\label{densities}
\end{figure}

\begin{figure}[!ht]
\centerline{\includegraphics[width=5.25cm,clip,angle=-90,bb=175 475 455 810]{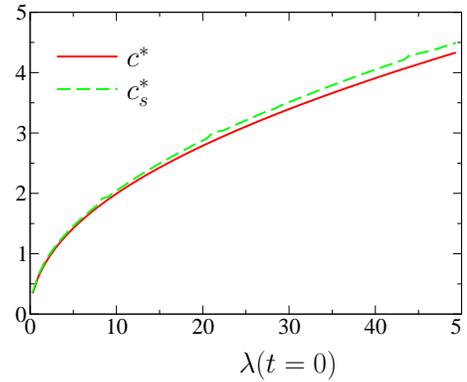}}
\caption{(Color online) Goldstone mode velocity $c^*$ and macroscopic sound
  velocity $c^*_s$ {\it vs.} $\lambda(t=0)$ for $d=2$ and $n_0(t=0)=0.2$.} 
\label{velocities}
\end{figure}

\begin{figure}[!ht]
\centerline{\includegraphics[width=5.25cm,clip,angle=-90,bb=175 475 455 810]{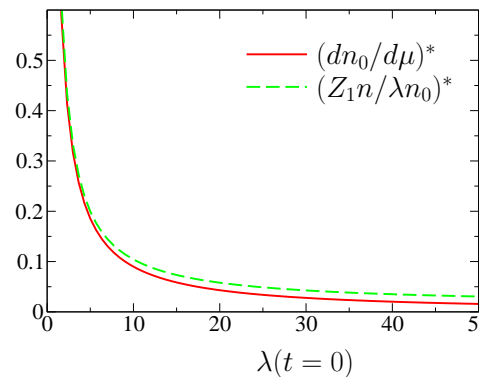}}
\caption{(Color online) $Z_1^*n^*/(\lambda^* n^*_0)$ and the condensate
  ``compressibility'' $(dn_0/d\mu)^*$ {\it vs.} $\lambda(t=0)$ for $d=2$ and
  $n_0(t=0)=0.2$.}  
\label{kappa}
\end{figure}

There are three important relations between 1PI vertices and thermodynamic
quantities that our results should fulfill, 
\bleq
 & n_s=n=-\dfrac{\partial\Omega}{\partial\mu}, \quad
c=c_s=\left(\dfrac{n}{m(dn/d\mu)}\right)^{1/2}, & \nonumber \\ &
\dfrac{Z_1 n}{\lambda n_0} = \dfrac{dn_0}{d\mu} . &
\label{equalities}
\eleq
The equality of the superfluid density $n_s$ and the density $n$ is a
consequence of Galilean invariance at zero temperature.  
Since the chemical potential appears only in the initial conditions for
the effective action $\Gamma[\phi]$, derivatives with respect to $\mu$ can be
numerically calculated by solving the flow equations for nearby values of
$\mu$. The equality between the Goldstone mode velocity $c$ and the macroscopic
sound velocity $c_s$ was proved in Ref.~\cite{Gavoret64}. The last relation
in (\ref{equalities}) is a consequence of gauge invariance \cite{Pistolesi04}.
The figures \ref{densities}, \ref{velocities} and \ref{kappa} show that
the symmetry constraints (\ref{equalities}), despite a good overall agreement,
are not strictly enforced in our approach. 
This can be ascribed to the choice of our infrared regulator
$R(\q^2)$ as well as the Ansatz (\ref{ansatz}) which are both incompatible with
Galilean invariance. 

Making use of (\ref{equalities}), we can rewrite the propagator $\bar
G(q)=-\bar\Gamma^{(2)-1}(q)$ as 
\bleq
\bar G_{22}(q) &=& - \frac{2mc^2n_0}{n} \frac{1}{\w^2+c^2\q^2} , \nonumber \\
\bar G_{12}(q) &=& \frac{mc^2}{n}\frac{dn_0}{d\mu} \frac{\w}{\w^2+c^2\q^2} ,
\nonumber \\ 
\bar G_{11}(q) &=& - \frac{1}{2\lambda n_0}  
\label{propa} 
\eleq
in the infrared limit. In (\ref{propa}), all quantities except $\lambda$ can
be evaluated at $k=0$. Because of the vanishing of $\lambda(k\to 0)$,
the longitudinal correlation function $\bar G_{11}$
diverges logarithmically in three dimensions and as $k^{-\eps}$ below
\cite{Nepomnyashchii75,Nepomnyashchii78,Popov79}. (see
table \ref{table}). The dependence on $q$ can be restored
by evaluating $\lambda$ at $k\sim\sqrt{\w^2+c^2\q^2}$. In
Ref.~\cite{Pistolesi04}, the equations (\ref{propa}) were obtained by imposing
the Ward identities due to gauge invariance and solving a one-loop RG equation
for the sole independent coupling in the limit $k\to 0$ \cite{note4}. 
By a detailed
analysis of the structure of the perturbation theory to higher order, it was
then argued that the equations (\ref{propa}) give the exact asymptotic
behavior. We believe that our RG approach, being intrinsically non-perturbative
\cite{Berges00}, gives further support to this claim.  

\section{Superfluidity without BEC ($d=1$)} 

In one-dimension, as a result of the emerging SO(2) symmetry, we find that the
long-distance physics is described by the classical O(2) model in $d+1=2$
dimensions \cite{Wetterich07}. We thus expect the system to be in 
the ``low-temperature'' phase of the Kosterlitz-Thouless phase transition. 
There is no BEC as the condensate density $n_0\sim k^{\eta^*}$ vanishes in the
thermodynamic limit $k\to 0$. However, the superfluid density $n_s=Zn_0$
remains finite. This phase is generally described as a Luttinger liquid (LL)
characterized by the Goldstone mode velocity $c^*$ and the LL parameter $K$
\cite{Haldane81,Cazalilla04}.  

It has been shown that the NPRG gives a good description of the classical O(2)
model \cite{Graeter95,Gersdorff01}. In particular the low-temperature phase is
characterized by an approximate line of fixed points where the beta function
becomes very small and the running of the renormalized order parameter
$\tilde n_0'$ (or, equivalently, the phase stiffness) very slow, which implies
a very large, although not strictly infinite, correlation length $\xi$. The
anomalous exponent $\eta$ depends on the (slowly running) order parameter
$\tilde n_0'$ and takes its largest value $\sim 1/4$ when the system crosses
over to the disordered regime ($k\sim\xi^{-1}$).

\begin{figure}[!ht]
\centerline{\includegraphics[width=7.5cm]{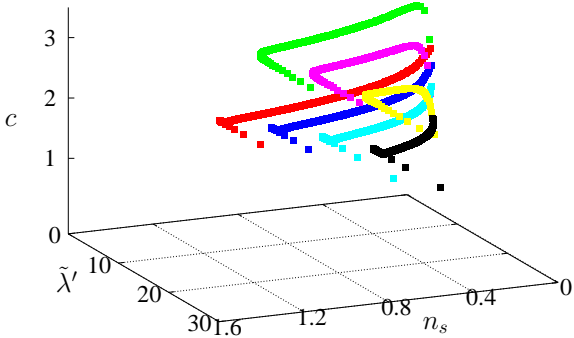}}
\caption{(Color online) RG trajectories $(n_s,\tilde\lambda',c)$ in one
  dimension for various initial conditions $n_0(t=0)$ and $\lambda(t=0)$. The
  points correspond to equal steps in $t$.} 
\label{flow1_1D}
\end{figure}

\begin{figure}[!ht]
\centerline{\includegraphics[width=6.5cm]{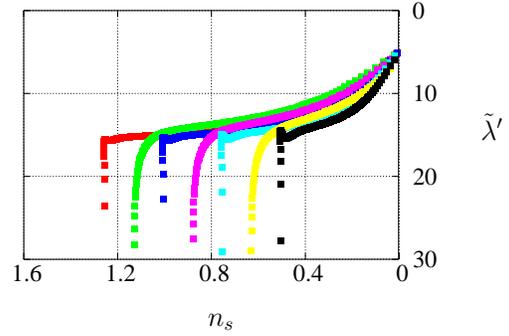}}
\caption{(Color online) Same as in Fig.~\ref{flow1_1D} but in the plane
  $(n_s,\tilde\lambda')$.} 
\label{flow2_1D}
\end{figure}

\begin{figure}[!ht]
\centerline{\includegraphics[width=6cm]{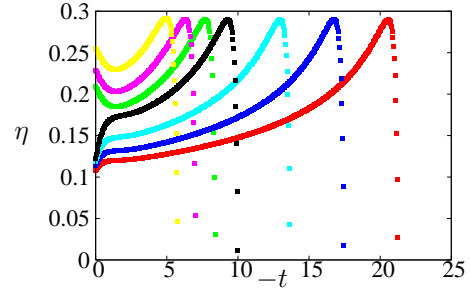}}
\caption{(Color online) Anomalous exponent $\eta$ {\it vs.} $-t$ for various
  initial conditions $n_0(t=0)$ and $\lambda(t=0)$.}
\label{eta_1D}
\end{figure}

By solving numerically the flow equations (\ref{rgeq}) in one dimension, we
have obtained very similar results. Figs.~\ref{flow1_1D} and \ref{flow2_1D}
show the flow trajectories in the space $(n_s,c,\tilde\lambda')$ for various
initial conditions $n_0(t=0)$ and $\lambda(t=0)$. The points correspond to
equal steps in $t$ so that very dense points indicate a very slow running. 
For a sufficiently small
ratio $\lambda(t=0)/n_0(t=0)$, we find that trajectories rapidly hit an
approximate plane of fixed points defined by $\tilde\lambda'\sim 15$, where
the running of the superfluid density $n_s$ and the Goldstone mode velocity
$c$ become very slow. As for the classical O(2) model, we infer from this
observation that the correlation length $\xi$ is extremely large for these
trajectories. For very long RG time $-t$ ($k\sim\xi^{-1}$), the system
eventually crosses over to the disordered regime. 
On the approximate plane of fixed points, the scale-dependent anomalous
exponent $\eta$ varies slowly about a value that depends both on $n_s$ and $c$
(Fig.~\ref{eta_1D}). It then reaches its maximum value $\eta_c\simeq 0.29$ -- 
to be compared with the exact exponent $\eta_c=1/4$ at the Kosterlitz-Thouless
transition of the classical O(2) model -- before rapidly dropping to zero once
$k\sim\xi^{-1}$.   
For $M\gg 1$, which corresponds to a large superfluid density $n_s$, we
obtain the analytic expression $\eta = 
mc/(2\pi n_s)$ from (\ref{rgeq}). This in turn
determines the LL parameter $K=1/(2\eta)=\pi n_s/(mc)$, which is the
expected value in a Galilean invariant system where $n_s=n$ \cite{Haldane81}. 

Thus for a sufficiently small ratio $\lambda(t=0)/n_s$ (or $c/n_s$), we obtain
a good picture of 
the Luttinger liquid behavior of the superfluid phase. When this ratio is too
large, the flow trajectory does not reach the approximate plane of fixed
points, and the system is in the ``high-temperature'' (disordered)  phase of
the classical O(2) model. This result is in contradiction with known results
in one dimension where the action (\ref{action})
corresponds to the exactly soluble Lieb-Liniger model \cite{Lieb63a,Lieb63b}.
This model is parameterized by the dimensionless parameter
$\gamma=m\lambda(t=0)/n$ and  
its low-energy description is a Luttinger liquid with a parameter
$K\equiv K(\gamma)$ varying in the interval $[1,\infty[$ as $\gamma$
decreases from infinity to zero \cite{Cazalilla04}. The limit $\gamma\to\infty$
($K=1$) corresponds to hard-core bosons. Thus the anomalous exponent
$\eta$ should take its highest value $1/(2K_{\rm min})=1/2$ for
$\gamma\to\infty$ rather than $1/4$ as predicted by our results. 
A possible explanation for the failure of our approach to correctly describe
the strong-coupling limit of the action (\ref{action}) in one dimension is the
derivative expansion used in the Ansatz (\ref{ansatz}) for $\Gamma[\phi]$.
Quite generally, the derivative expansion is known to work best
when $\eta$ is small \cite{Berges00,Delamotte07}.

\section{Conclusion} 

The NPRG technique discussed in this Letter provides an efficient method to
control the infrared divergences appearing in the perturbation theory of
zero-temperature Bose systems. It extends the approach of
Ref.~\cite{Wetterich07} and reproduce the results obtained earlier by a
field-theoretical RG approach combined with the implementation of Ward
identities due to gauge invariance \cite{Castellani97,Pistolesi04}. The
non-trivial infrared behavior in  dimensions $1<d\leq 3$, characterized by the
divergence of the longitudinal 
correlation function and the vanishing of the anomalous self-energy
$\Sigma_{\rm an}(q\to 0)$, turns out to be related to the emergence of a
space-time SO($d$+1) symmetry at low energy. This implies a close link between
the superfluid phase and the Goldstone regime of the classical O(2) model in
$d+1$ dimension \cite{Wetterich07}. 
 
Our approach also describes one-dimensional systems where superfluidity exists
without BEC in the thermodynamic limit. The superfluid phase exhibits a
Luttinger-liquid behavior that is well captured by the NPRG
approach for weak interactions. Although our results, based on a derivative
expansion of the effective action $\Gamma[\phi]$, break down at strong
coupling, they might be improved by a more refined treatment of the momentum
dependence of the vertices \cite{note5}.

An important feature of the NPRG is that it not only yields the infrared
behavior of correlation functions but can also compute propagators 
in terms of the parameters of a microscopic
model. It thus provides an efficient tool for the explicit calculation of
physical quantities beyond the Bogoliubov theory while satisfying basic
requirements such as the Hugenholtz-Pines theorem as well as yielding the
correct infrared behavior, a task that has been known
to be difficult in interacting boson systems \cite{Hohenberg65,note6}.

\acknowledgments

We thank the Mathematics department of Imperial College for hospitality.
ND is grateful to B. Delamotte, D. Mouhanna, M. Tissier, L. Canet and
N. Wschebor for enlighting discussions and/or correspondence on
the NPRG. KS thanks CAMCS for support.


\begin{thebibliography}{10}

\bibitem{Bogoliubov47}
N.~N. Bogoliubov, J. Phys. USSR {\bf 11},  23  (1947).

\bibitem{Hohenberg65}
P.~C. Hohenberg and P.~C. Martin, Ann. Phys. (N.Y.) {\bf 34},  291  (1965).

\bibitem{Hugenholtz59}
N. Hugenholtz and D. Pines, Phys. Rev. {\bf 116},  489  (1959).

\bibitem{Beliaev58b}
S.~T. Beliaev, Sov. Phys. JETP {\bf 7},  289  (1958); {\bf 7},  299  (1958).

\bibitem{Gavoret64}
J. Gavoret and P. Nozi\`eres, Ann. Phys. (N.Y.) {\bf 28},  349  (1964).

\bibitem{Nepomnyashchii75}
A.~A. Nepomnyashchii and Y.~A. Nepomnyashchii, JETP Lett. {\bf 21},  1  (1975).

\bibitem{Nepomnyashchii78}
Y.~A. Nepomnyashchii and A.~A. Nepomnyashchii, Sov. Phys. JETP {\bf 48},  493
  (1978).

\bibitem{Nepomnyashchii83}
Y.~A. Nepomnyashchii, Sov. Phys. JETP {\bf 58},  722  (1983).

\bibitem{Popov79} V.~N. Popov and A.~V. Seredniakov, Sov. Phys. JETP {\bf 50},
  1 (1979). 

\bibitem{note2}
The normal ($\Sigma_{\rm n}$) and anomalous ($\Sigma_{\rm an}$) self-energies
are commonly denoted by $\Sigma_{11}$ and 
  $\Sigma_{12}$, where the index $i=1,2$ refers to the two components of the
  field $\Psi=(\psi,\psi^*)^T$. Since the index $i$ bears a different meaning
  in our approach, we refrain from using the common notation.

\bibitem{Patasinskij74}
A.~Z. Patasinskij and V.~L. Pokrovskij, Sov. Phys. JETP {\bf 37},  733  (1973).

\bibitem{Weichman88}
P.~B. Weichman, Phys. Rev. B {\bf 38},  8739  (1988).

\bibitem{Benfatto94}
G. Benfatto,  in {\em Constructive results in field theory, statistical
  mechanics, and condensed matter physics}, edited by V. Rivasseau (Springer
  Verlag, New York, 1994).

\bibitem{Castellani97}
C. Castellani, C.~D. Castro, F. Pistolesi, and G.~C. Strinati, Phys. Rev. Lett.
  {\bf 78},  1612  (1997).

\bibitem{Pistolesi04}
F. Pistolesi, C. Castellani, C.~D. Castro, and G.~C. Strinati, Phys. Rev. B
  {\bf 69},  024513  (2004).

\bibitem{Bloch07} For a recent review on ultracold gases, see for instance 
  I. Bloch, J. Dalibard and W. Zwerger, arXiv:0704.3011 (unpublished). 

\bibitem{Wetterich91}
C. Wetterich, Nucl. Phys. B {\bf 352},  529  (1991).

\bibitem{Wetterich93}
C. Wetterich, Phys. Lett. B {\bf 301},  90  (1993).

\bibitem{Berges00}
For a review on the NPRG, see 
J. Berges, N. Tetradis, and C. Wetterich, Phys. Rep. {\bf 363},  223  (2000).

\bibitem{Delamotte07}
For a pedagogical introduction to the NPRG, see 
B. Delamotte, arXiv:cond-mat/0702365 (unpublished).

\bibitem{Andersen99} J.~O. Andersen and M. Strickland, Phys. Rev. A
  {\bf 60}, 1442 (1999).

\bibitem{Wetterich07} C. Wetterich, arXiv:0705.1661 (unpublished).

\bibitem{note1}
The most general derivative expansion to order ${\cal O}(\partial^2)$ would
  include the terms $Y\nablabf n \nablabf n$ and $V' \dtau n \dtau n$. These
  are not expected to play an important role and are neglected.

\bibitem{note3} $\bar\Gamma^{(2)}_{22}(q=0)=-\mu+\Sigma_{\rm n}(0)-\Sigma_{\rm
  an}(0)=0$ is an exact statement of the Hugenholtz-Pines theorem. $\Sigma_{\rm
  n}$ and $\Sigma_{\rm an}$ denote the normal and anomalous self-energies
  \cite{note2}.   

\bibitem{Litim00}
D. Litim, Phys. Lett. B {\bf 486},  92  (2000).

\bibitem{note4}
If one neglects finite renormalization corrections and use (\ref{equalities})
  as well as $2\eta^*_1+\eta^*_2=-2$, one is left with only one independent
  running coupling (e.g. $\lambda$ or $2\lambda n_0$).

\bibitem{Haldane81}
F.~D.~M. Haldane, Phys. Rev. Lett. {\bf 47},  1840  (1981).

\bibitem{Cazalilla04}
M.~A. Cazalilla, J. Phys. B {\bf 37},  S1  (2004).

\bibitem{Graeter95}
M. Gr\"ater and C. Wetterich, Phys. Rev. Lett. {\bf 75},  378  (1995).

\bibitem{Gersdorff01}
G.~V. Gersdorff and C. Wetterich, Phys. Rev. B {\bf 64},  054513  (2001).

\bibitem{Lieb63a}
E.~H. Lieb and W. Liniger, Phys. Rev. {\bf 130},  1605  (1963).

\bibitem{Lieb63b}
E.~H. Lieb, Phys. Rev. {\bf 130},  1616  (1963).

\bibitem{note5}
For an NPRG scheme that goes beyond the derivative expansion, see J.-P.
  Blaizot, R. M\'endez-Galain and N. Wschebor, Phys. Lett. B {\bf 632}, 571
  (2006); A. Sinner, N. Hasselmann, P. Kopietz, arXiv:0707.4110 (unpublished).

\bibitem{note6}
For a review of approximations beyond the Bogoliubov theory, see for instance
  H. Shi and A. Griffin, Phys. Rep. {\bf 304}, 1 (1998),
  J.~O. Andersen, Rev. Mod. Phys. {\bf 76}, 599 (2004). 

\end{thebibliography}
\end{document}